\documentclass[journal]{IEEEtran}
\usepackage{amsmath,amssymb,amsfonts,amsthm}
\usepackage{graphicx}
\usepackage{epstopdf}
\usepackage{subfigure}
\usepackage[justification=centering]{caption}
\usepackage{color}
\usepackage{float}
\usepackage[numbers, comma, sort&compress]{natbib}

\begin{document}
\title{Optimization Techniques in Reconfigurable Intelligent Surface Aided Networks}

\author{
	\IEEEauthorblockN{Biqian Feng, Junyuan Gao, Yongpeng Wu, Wenjun Zhang, Xiang-Gen Xia, and Chengshan Xiao}

	\thanks{B. Feng, J. Gao, Y. Wu, W. Zhang are with the Department of Electronic Engineering, Shanghai Jiao Tong University, Minhang 200240, China (e-mail: fengbiqian@sjtu.edu.cn; sunflower0515@sjtu.edu.cn; yongpeng.wu@sjtu.edu.cn; zhangwenjun@sjtu.edu.cn) (Corresponding author: Yongpeng Wu).}
	
	\thanks{X.-G. Xia is with the Department of Electrical and Computer Engineering, University of Delaware, Newark, DE 19716, USA. (e-mail: xxia@ee.udel.edu).}
	
	\thanks{C. Xiao is with the Department of Electrical and Computer Engineering, Lehigh University, Bethlehem, PA 18015 USA (e-mail: xiaoc@lehigh.edu).}
}

\maketitle

\centerline{\textbf{\Large Abstract}}

Reconfigurable intelligent surface (RIS)-aided networks have been investigated for the purpose of improving the system performance. However, the introduced unit modulus phase shifts and coupling characteristic bring enormous challenges to the optimization in the RIS-aided networks. Many efforts have been made to jointly optimize phase shift vector and other parameters. This article intends to survey the latest research results about the optimization in RIS-aided networks. A taxonomy is devised to categorize the existing literatures based on optimization types, phase shift form, and decoupling methods. Furthermore, in alternating optimization framework, we introduce in detail how to exploit the aforementioned technologies flexibly. It is known that most works could not guarantee a stationary point. To overcome this problem, we propose a unified framework for the optimization problem of RIS-aided networks with continuous phase shifts to find a stationary point. Finally, key challenges are outlined to provide guidelines for the domain researchers and designers to explore more efficient optimization frameworks, and then open issues are discussed.

\vspace{12 pt}
\centerline{\textbf{\Large Introduction}}
The explosive growth of mobile devices, the rapidly increasing demand on the broadband and high-rate communication services result in the increasingly severe spectrum scarcity problem. To enhance the communication performance, a novel concept of reconfigurable intelligent surface (RIS) has been introduced as a promising technique due to its capability of achieving high spectral efficiency and energy efficiency. A RIS comprises of a large number of low-cost passive antennas that can smartly reflect the impinging electromagnetic waves for performance enhancement. In RIS-aided networks, a base station (BS) firstly sends control signals to a RIS controller so as to jointly optimize the properties of incident waves and improve the quality of service. Then since BS sends signals to users and RIS at the same time, each user in general receives the superposed (desired as well as interference) signal from both BS-user link (direct) and BS-RIS-user (reflected) \cite{Wu SDR}. Therefore, this provides more flexibility for the design at the BS to upgrade the system performance.

How to design the parameters in RIS-aided networks has become a hot topic for recent and future wireless communications. 
The joint design is usually formulated as an optimization problem. 
In many works, the original optimization problem faces two main issues: i) the problem is NP-hard in essence due to the unit modulus phase shifts; ii) there always exists the coupling between phase shifts and other parameters (e.g. beamforming), which is caused by BS-RIS-user link in both objective function and constraints. Since most optimization problems are NP-hard in the RIS-aided networks, the globally optimal solutions are impossible to obtain in general. Therefore, instead of finding globally optimal solution, locally optimal solution (e.g. stationary point) or other efficient suboptimal heuristics are usually applied for practical implementations. Fortunately, even if the optimal solution is not reached, many works have verified that the performance of the system can be greatly improved \cite{Huang}. When the phase shift vector is given, the original problem reduces to a conventional communication problem without RIS, which has been investigated for decades and for which compelling mechanisms exist. Motivated by this idea, alternative optimization (AO) framework is often employed to upgrade the system performance constantly in the most works, which results in a low complexity but performance loss due to no guarantee of a stationary point \cite{Schober IA}. To further improve performance, there is a strong interest in computing locally optimal solution.

Successive convex approximation (SCA) technique is a powerful and general tool to optimize nonconvex problem.  
Many SCA techniques are proposed, such as multi blocks SCA, parallel SCA \cite{Scutari}, and constrained stochastic SCA (CSSCA) \cite{Liu}. Motivated by CSSCA, we provide a unified framework for the RIS-aided networks to find a stationary point. In this framework, we take strongly convex functions to approximate both the objective function and constraints. In many algorithms \cite{Schober IA,Scutari}, a tight convex bound with respect to all variables is indispensable in constraints, which increases the difficulty of implementation. It is worth noting that the proposed framework only requires strong convexity, gradient consistency, and value consistency, but it does not require a global bound, which increases the flexibility of implementation. Recalling the second issue aforementioned above, the nonconvexity of coupling in constraints becomes easy to handle. Furthermore, motivated by distributed implementation in parallel SCA \cite{Scutari}, we could decompose the approximate problem into distributively solvable subproblems which are then coordinated by a high-level master problem.

The contributions of our work are summarized as follows. This article intends to survey the latest research results about the optimization in RIS-aided networks. Firstly, we categorize most of the optimization problems into four basic optimization types. Then, we emphasize two key techniques in optimization problems, namely phase shift form and decoupling methods. Furthermore, in classic AO framework, we introduce how to exploit the aforementioned technologies flexibly in detail. Motivated by \cite{Scutari,Liu}, we propose a unified framework to find a stationary point without any requirement for bounds. Therefore, it is highly flexible to implement. Finally, we clarify the challenges and discuss the open issues as the future research directions.

\begin{figure*}[ht]
	\centering
	\fbox{\includegraphics[scale=0.65]{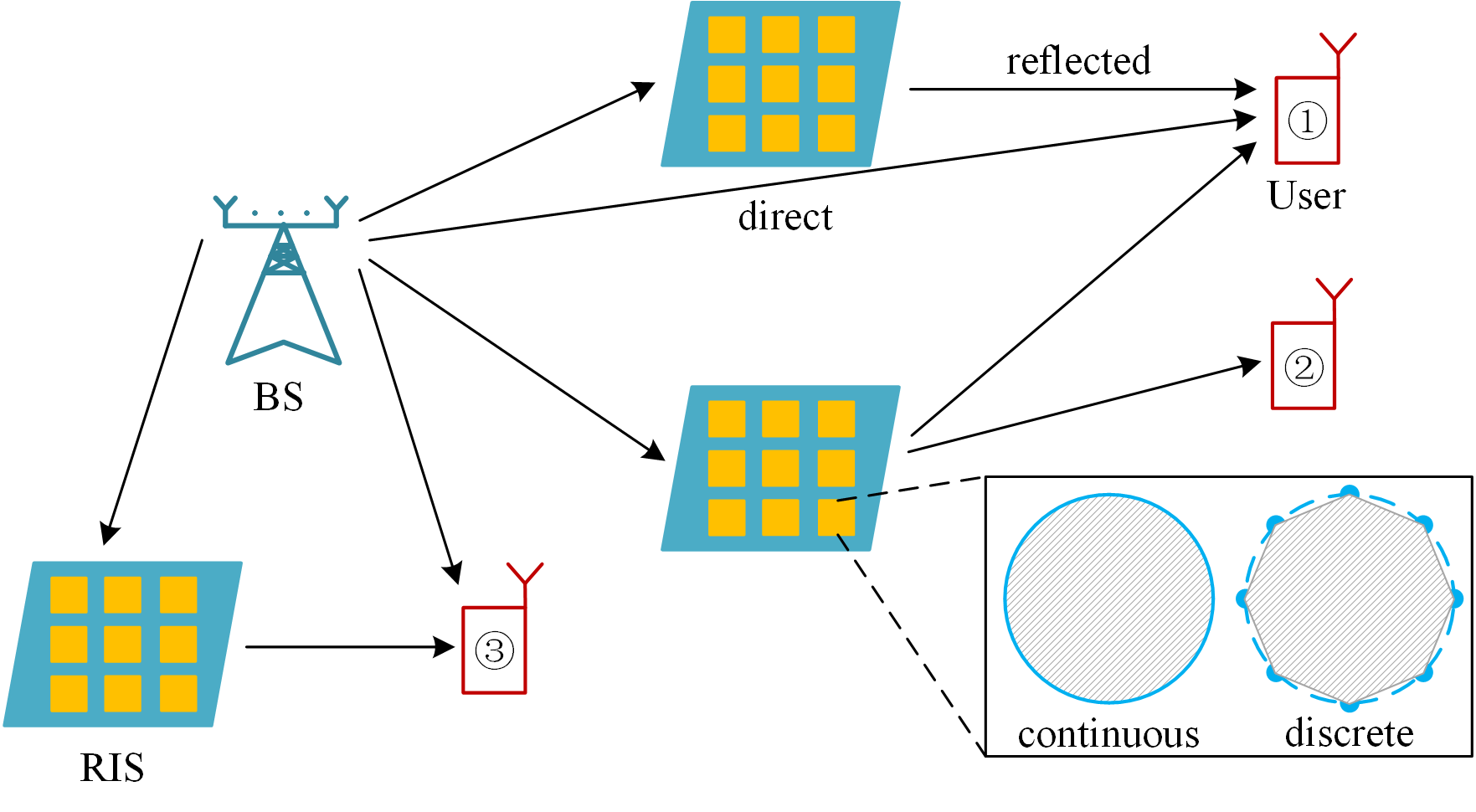}}
	\caption{Illustration of RIS-aided networks. User 1 is assisted by multiple RISs while user 2 and user 3 are assisted by a single RIS, but the direct link for user 2 is blocked severely.}
	\label{Systems}
\end{figure*}
\vspace{12 pt}
\centerline{\textbf{\Large Optimization Types and Applications}}
As illustrated in Fig. \ref{Systems}, the architecture of RIS-aided wireless communication networks consists of three basic components: i) BS; ii) RIS: single RIS or multiple RISs; iii) the target users. 
In this section, we categorize the optimization problem in RIS-aided networks into the following four basic problems according to the coupling between phase shift vector $\boldsymbol \phi$ and other system parameters $\mathbf w$:

\begin{itemize}
	\item \textbf{Problem 1:} $\boldsymbol \phi$ and $\mathbf w$ are coupled in the objective function while they are independent in constraints. For example, in \cite{Schober manifold}, the system aims to maximize achievable secrecy rate when there is no other constraint except for the power consumption of beamforming and artificial noise in downlink transmission. Since the achievable secrecy rate is related to the environment (e.g. phase shifts), beamforming, and artificial noise, both $\boldsymbol \phi$ and $\mathbf w$ are in objective function at the same time.
	
	\item \textbf{Problem 2:} Only $\mathbf w$ exists in the objective function while $\boldsymbol \phi$ and $\mathbf w$ are coupled in constraints. For example, in \cite{Wu SDR}, power consumption is adopted as a metric to evaluate the system performance while signal-to-interference-plus-noise ratio (SINR) constraints must be met at all users. Similarly, $\boldsymbol \phi$ and $\mathbf w$ are in SINR constraints because the achievable secrecy rate is related to the environment and beamforming.
	
	\item \textbf{Problem 3:} Only $\boldsymbol \phi$ exists in the objective function while $\boldsymbol \phi$ and $\mathbf w$ are coupled in constraints. For example, in multiple RISs-aided networks, it is often not energy efficient to turn on all the RISs. An interesting work is to minimize the number of turning-on RISs with all target users' requirements satisfied.
	
	\item \textbf{Problem 4:} $\boldsymbol \phi$ and $\mathbf w$ are coupled both in objective function and constraints. For example, in \cite{Zhaohui}, multiple RISs are spatially distributed to serve wireless users and the problem is posed as a joint optimization problem of beamforming and RIS control, whose goal is to maximize the energy efficiency under minimum rate constraints of the users. It can be seen that $\boldsymbol \phi$ and $\mathbf w$ are in the formulas of  energy efficiency and achievable rate simultaneously.
\end{itemize}

The main characteristic of Problem 2 and Problem 3 is that when the variable in the objective is fixed, they reduce to feasibility check problems. So, to ensure convergence, a new objective function needs to be reconstructed \cite{Wu SDR,Huang}. The main difference between Problem 1 and Problem 4 is that the nonconvex constraints in Problem 4 are more challenging to handle due to the coupling, which results in the hardness of finding the stationary point.

\vspace{12 pt}
\centerline{\textbf{\Large Phase Shifts at the RIS}}

In this section, we first provide the phase shift vector types for the RIS in practice, and then present some methods for overcoming the first issue in Introduction from different perspectives.

\vspace{6 pt}
\centerline{\textbf{Phase Shift Types}}

Let $\boldsymbol{\phi}\triangleq\left(e^{j\theta_1},e^{j\theta_2},\cdots,e^{j\theta_N}\right)^T$ be the reflection-coefficient vector of the RIS with $N$ passive beamforming elements, and $\theta_n$ and $\phi_n\triangleq e^{j\theta_n}$ denote the $n$-th phase shift of the RIS.

\begin{itemize}
	\item \textbf{Type 1:} $\theta_n\in[0,2\pi)$. The continuous phase shift vector is adopted widely in many works \cite{Wu SDR,Huang}, because it is easier to be handled than discrete phase shifts in Type 2.
	
	\item \textbf{Type 2:} $\theta_n\in\{0, \Delta\theta, \cdots, (2^b-1)\Delta\theta\}$, where $b$ denotes the number of bits used to indicate the number of phase shift levels and $\Delta\theta=2\pi/2^b$. For ease of practical implementation, the phase shift at each element of the RIS takes only a finite number of discrete values \cite{Wu Discrete}.
\end{itemize}

Note that types 1 and 2 are both the ideal phase shift models assuming full signal reflection by each of its elements regardless of the phase shift, which, however, is practically difficult to realize. In contrast, based on the property of narrowband and wideband channels, two practical phase shift models $\beta_n\left(\theta_n\right)e^{j\theta_n}$ with different amplitude forms are proposed to capture the phase-dependent amplitude variation in the element-wise reflection design in references \cite{Abeywickrama Practical, Li Practical}.


\vspace{6 pt}
\centerline{\textbf{Schemes for Continuous Phase Shifts}}

For the purpose of summarizing the different schemes more clearly, some equivalent forms of the phase shift vector are shown in three levels, namely phase $\theta_n$, complex exponential $\phi_n$, and the composition of phase shift vector.
\begin{itemize}
	\item \textbf{1st form:} $\theta_n\in[0,2\pi)$: We treat the phase shift $\theta_n$ as an optimization variable directly, which implies the constraint is convex. In many works \cite{Huang,Schober Large}, there are no more constraints containing $\theta_n$, and thus the problem becomes an unconstrained programming problem so that many classic techniques, such as gradient-based algorithms, are exploited to monotonically decrease its objective function, eventually converging to a stationary point when other variables are given.
	
	\item \textbf{2nd form:} 1) $|\phi_n|=1$; 2.1) $|\phi_n|\geq 1, |\phi_n|\leq 1$ or $N-\|\boldsymbol\phi\|\geq 0, |\phi_n|\leq 1$; 2.2) $u_n=\phi_n, |u_n|=1$; 3) $|\phi_n|\leq 1$ + Projection. It is clearly seen from Fig. \ref{Systems} that each element lies on a circle (blue), so the constraint is nonconvex. At this level, the ideas of most works can be classified into three categories. The first one is that the algorithm is suitable for the whole domain and then we solve it in the feasible region. In majorization-minimization (MM) algorithm \cite{Huang}, the authors construct an upper-bound function in the domain, and get an optimal solution of approximate problem in the feasible region at each iteration. The second one is a penalty method, in which the penalty terms for the constraint violations (e.g. nonconvex constraints in 2.1) and $u_n=\phi_n$ in 2.2)) are multiplied by a positive coefficient. By making this coefficient larger, we penalize constraint violations more severely, thereby forcing the minimizer of the penalty function closer to the feasible region for the constrained problem \cite{You Penalty,Zhaohui,Ming-Min PDD}. The last one is a low-complexity scheme which relaxes $\phi_n$ to its convex hull (shaded area in Fig. \ref{Systems}), then projects the solution to the unit circle \cite{Gui KKT}.
	
	\item \textbf{3rd form:} $\mathbf V=\boldsymbol{\phi}\boldsymbol{\phi}^H$: 1) Stiefel manifold; 2) $\operatorname{Rank}\left(\mathbf V\right)=1 (\|\mathbf V\|_*-\|\mathbf V\|_2\leq 0),\operatorname{diag}\left(\mathbf V\right)=1,\mathbf V\succeq\mathbf 0$. At this level, the constraint defines a Stiefel manifold, and the constraint is automatically satisfied when $\boldsymbol{\phi}$ is optimized over the Stiefel manifold. Based on this, we can exploit the optimization approaches designed for the Euclidean space to tackle manifold optimization problems \cite{Schober manifold}. On the other hand, in 2), since the rank-one constraint is nonconvex, there exist two methods to ``remove'' this constraint, namely penalty method \cite{Schober IA} and semidefinite relaxation (SDR) \cite{Wu SDR}. The main idea of penalty method is the same as the 2nd form, while SDR is applied to relax the constraint, then a rank-one solution is constructed from the optimal higher-rank solution with Gaussian random vector in 2).
\end{itemize}

It can be seen that penalty method is usually exploited in the 2nd and 3rd forms. However, the 3rd form lifts the problem dimension (i.e., the number of variables) from $N$ to $N^2$, which drastically increases the memory burden and computational cost when the problem is large-scale. The 2nd form has two distinct advantages, i) low-complexity at each iteration: reference \cite{You Penalty} adopts GEMM method and the computational complexity of each iteration is $\mathcal{O}\left(N^2\right)$, while semidefinite programming (SDP) is used to the 3rd form with the computational complexity $\mathcal{O}\left(N^{3.5}\right)$; ii) high flexibility: it is highly flexible for the 2nd form to split the phase shifts into smaller blocks when massive RIS elements are used.

\vspace{6 pt}
\centerline{\textbf{Schemes for Discrete Phase Shifts}}
\begin{itemize}	
	\item \textbf{4th form:} $\boldsymbol{\phi}_n=\left[0, \Delta\theta, \cdots, (2^b-1)\Delta\theta\right]^T \mathbf x$ with $\mathbf x$ being binary vector. Similarly, $\cos(\phi_n)$ and $\sin(\phi_n)$ can also be treated as a function of $\mathbf x$. In this way, since the problem converts the optimization variables $\boldsymbol{\phi}_n,\cos(\phi_n),\sin(\phi_n)$ into a unified binary vector $\mathbf x$, the problem is simplified a lot. Especially, the quadratic function can be transformed into an integer linear program, for which the globally optimal solution can be obtained by applying the branch-and-bound (BnB) method \cite{Wu Discrete}.
	
	\item \textbf{5th form:} $\theta_n=2\pi m/2^b, m\in\mathbb Z$. A low-complexity successive refinement (SR) algorithm where the optimal discrete phase shifts of different elements at the RIS are determined one by one in an iterative manner with those of the others being fixed \cite{Wu Discrete}.
	
	\item \textbf{6th form:} $|\phi_n|\geq 1, \phi_n\in \mathcal{U}$, where $\mathcal{U}$ is the convex hull of the feasible set (shaded area in Fig. \ref{Systems}). The main advantage of this form is that all the constraints are convex with a nice geometry structure for the set $\mathcal{U}$, which is similar to the 2nd form in continuous phase shifts. Since it transforms integer programming problem into continuous optimization problem, some methods in the 2nd form can be exploited for discrete phase shift optimization problem \cite{You Penalty}.
	
	\item \textbf{7th form:} $\mathbf u_n=\boldsymbol{\phi}_n, \mathbf u_n=2\pi m/2^b, m\in\mathbb Z$. By relaxing $\mathbf u_n=\boldsymbol{\phi}_n$ with penalty methods, $\boldsymbol{\phi}_n$ is not directly related to discrete phase shift set. In \cite{Ming-Min PDD}, an efficient penalty dual decomposition (PDD)-based algorithm is proposed, where the RIS phase shifts are updated in parallel to reduce the computational time.
	
\end{itemize}

\vspace{6 pt}
In practical phase shift models, the control of reflection coefficients is much more difficult due to the existence of the amplitude $\beta_n\left(\theta_n\right)$, and the difficulties mainly include reflection coefficient type (similar to phase shift type in ideal models) and the optimization in subproblems. Many schemes for ideal models lose efficacy. To alleviate the issue of designing reflection coefficients, two schemes have been proposed in references \cite{Abeywickrama Practical,Li Practical}: i) update only one element of RIS with fixed other elements; ii) introduce $\mathbf u_n=\boldsymbol \phi_n$, relax it with penalty methods, and update all phase shifts in parallel. However, subproblems are still complex with the existence of $\beta_n\left(\theta_n\right)$, and we can resort to the approximate solution.

\vspace{12 pt}
\centerline{\textbf{\Large Coupling in Constraints}}

As shown in Fig. \ref{Systems}, the coupling between $\boldsymbol{\phi}$ and $\mathbf w$ is inevitable because the fading channel of BS-RIS-Users link is $\mathbf h\operatorname{diag}\left(\boldsymbol{\phi}\right)\mathbf G$, where $\mathbf h$ and $\mathbf G$ represent RIS-User and BS-RIS channels, respectively. Therefore, how to address the coupling between $\boldsymbol{\phi}$ and $\mathbf w$ is always a key problem. In this section, we provide three methods in practice and discuss the effects and challenges.

\textbf{AO:} Since the other blocks are fixed in AO, the coupling will disappear automatically, which is an important factor in the widespread use of the AO framework. For example, in \cite{Huang}, given beamforming, the original problem becomes a phase shift optimization problem.

\textbf{Convex subset:} Replacing nonconvex constraints with convex subset is a common approach to alleviate the problem of nonconvex constraints. In practice, interior-point method, MM, and SCA \cite{Scutari} are usually based on this idea. It is worth pointing out that the generated solution at each iteration must satisfy the constraints. Instead of decoupling $\boldsymbol{\phi}$ and $\mathbf w$ in AO, the joint convexity with respect to all variables is a must in this method. For example, reference \cite{Schober IA} takes the jointly convex upper-bounds of the constraints with respect to beamforming and phase shifts.

\textbf{Convex approximation:} Although convex subset provides a great performance to alleviate the coupling, the jointly convex bound is not always easy to construct. Motivated by the reference \cite{Liu}, we aim to develop a strong convex approximation for constraints without any requirement for bounds, and the proposed algorithm guarantees a stationary point (please see the following section: A Unified Framework).

\vspace{6 pt}
\centerline{\textbf{Effects and Challenges}}
Recall three forms in continuous phase shifts. In the 2nd and 3rd forms, there exists no convex subset except a singleton. Specifically, in the 2nd form, as can be seen clearly from Fig. \ref{Systems}, there is no convex subset on the circle except a singleton because given any two points, the whole line segment that joins them is not on the circle. In the 3rd form, assuming $\mathbf V_1$ and $\mathbf V_2$ are in the convex subset of $\{\mathbf V\mid\mathbf V=\phi\phi^H\}$, it can be easily proved that $\mathbf V_1=\mathbf V_2$. Thus the subset is also a singleton, which causes the Slater's constraint qualification not to be satisfied. Therefore, convex subset-based algorithms are not readily to cope with the nonconvex constraints directly. To overcome this problem, some other methods, such as penalty, could be used to ``remove'' this troublesome constraint. 

\vspace{12 pt}
\centerline{\textbf{\Large Classic AO Framework}}
The problems formulated in many works are difficult to tackle mainly due to the nonconvex objective function, the unit modulus phase shift vector, and the coupled constraint sets. As we can see, the RIS-aided channel reduces to the conventional channel model with the phase shifts fixed. Based on this, all optimization variables are typically divided into two parts.
The AO framework aims to optimize the conventional variables and the phase shift vector alternatively with the other one fixed. The architecture of AO framework consists of three main steps:
\begin{itemize}
	\item \textbf{Step 1:} Preprocessing. To circumvent the intractable original function, we resort some classic techniques to convert it into a new approximation problem. For example,  a challenging spectral efficiency maximization problem is converted into a mean-squared error (MSE) minimization problem by Weighted Minimum MSE (WMMSE) \cite{You Penalty}. Lagrangian dual transform technique moves SINR to the outside of logarithm in achievable rate formula \cite{Huayan FP}. Penalty method moves the constraints to the objective function \cite{Schober IA}. In the most cases, they are equivalent in the sense that the optimal solutions of them are equal. It is worth noting that the stationary point of the new approximate problem may not be that of the original problem.
	
	\item \textbf{Step 2:} Optimize the conventional variables $\mathbf w$ with the phase shift vector $\boldsymbol{\phi}$ given. Many techniques have been investigated for decades without RIS, e.g. zero-forcing (ZF), minimum mean squared error (MMSE), maximum-ratio-transmission (MRT), etc.
	
	\item \textbf{Step 3:} Optimize the phase shift vector $\boldsymbol{\phi}$ with the conventional variables $\mathbf w$ given. Some techniques have been described in the previous sections.
\end{itemize}
Besides, AO framework can also be exploited for the subproblem in Step 2 and Step 3 \cite{Huayan FP}. In order to understand how to incorporate the above techniques into a specific scenario, we will have an analysis of some examples in this section.

In \cite{Wu SDR} and \cite{Wu Discrete}, the authors minimize the total transmit power at the BS by jointly optimizing the transmit beamforming at the BS and phase shifts at the RIS, subject to individual SINR constraints at all users (\textbf{Problem 2}). The resulting problem is tackled by AO framework in order to obtain a practical optimization method, at the price of sacrificing optimality. \textbf{Step 2:} The well-known MRT technique is used in single user system, while MMSE is taken to cope with the multiuser interference in multiuser system. \textbf{Step 3:} In continuous phase shift system, by taking the \textbf{3rd form} of continuous phase shifts and removing the rank constraint, SDR is taken to solve the phase shift problem,
while branch-and-bound method is applied to obtain globally optimal solution (\textbf{4th form}) and SR algorithm is exploited to obtain a suboptimal solution (\textbf{5th form}) in discrete phase shift system.

In AO framework, there is no requirement for handling with coupling since it disappears automatically, which enjoys low-complexity but performance loss. Inspired by these examples, we conclude that when a new optimization problem emerges, the following three aspects can be considered in turn: optimization type, conventional techniques for conventional variables, and phase shift forms.

\vspace{12 pt}
\centerline{\textbf{\Large Taxonomy}}
By summarizing the works in [1-3,6-15], Table~\ref{Categories} is given to show the taxonomy of optimization techniques. The taxonomy proposed in this article is based on optimization type, phase shift form, and decoupling.
\begin{table}
	\centering
	\fbox{\includegraphics[scale=0.65]{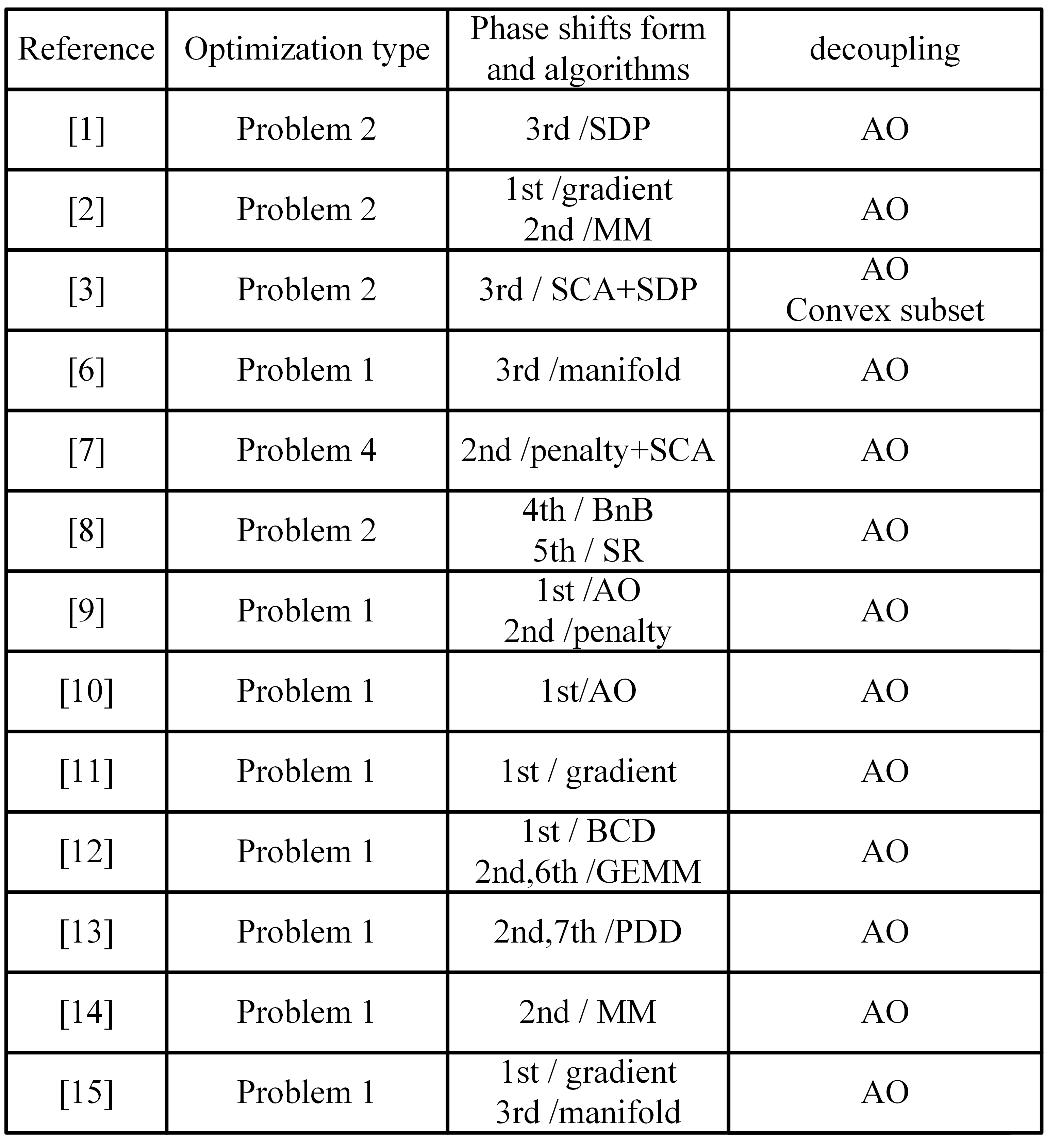}}
	\caption{Taxonomy of optimization techniques}
	\label{Categories}
\end{table}

\begin{figure*}[ht]
	\centering
	\fbox{\includegraphics[scale=0.65]{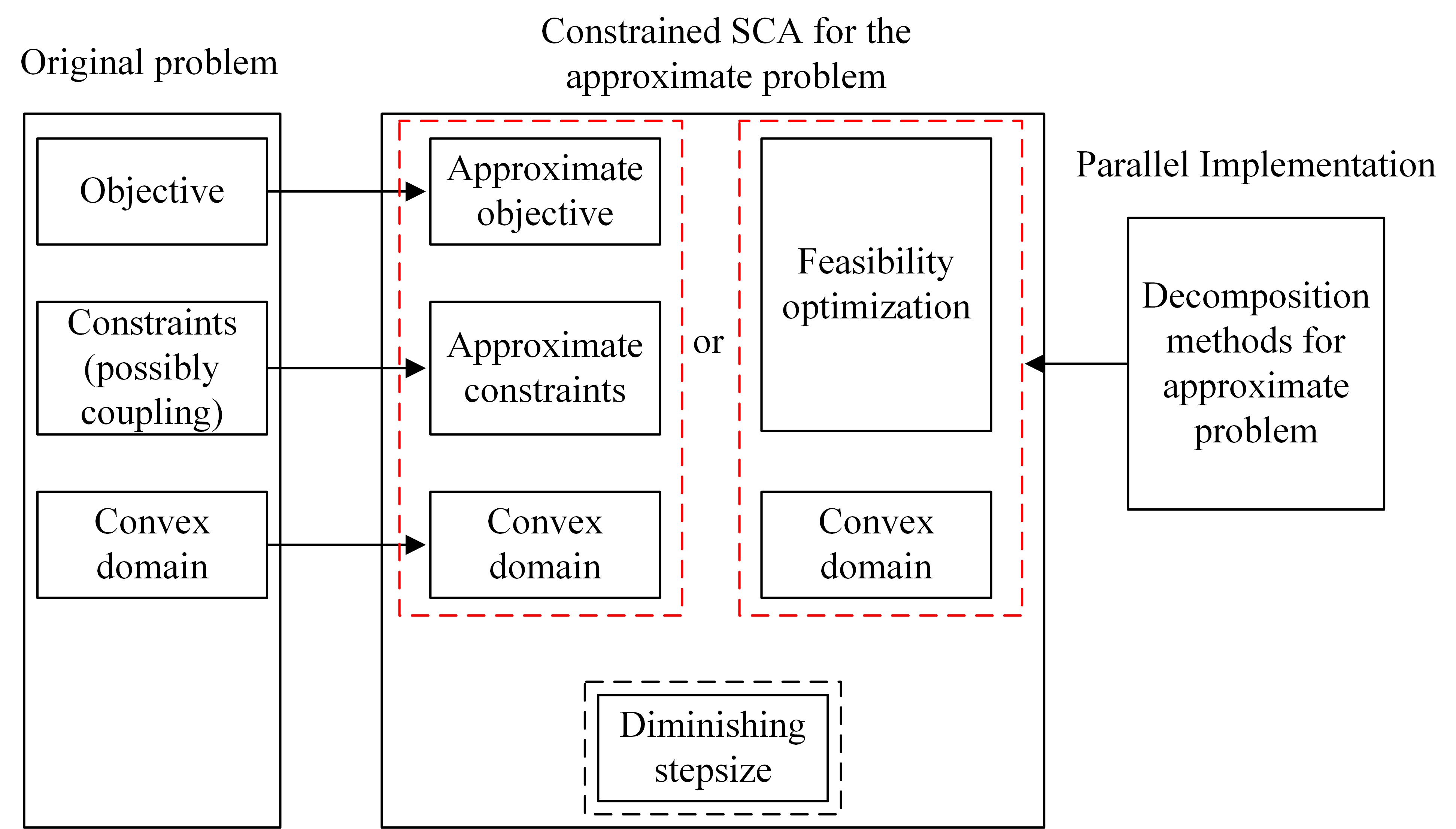}}
	\caption{A unified framework. All nonconvex function is approximated by strongly convex function with  gradient consistency and value consistency. The red dashed boxes represent strongly convex/feasibility optimization problems.}
	\label{Unified_Framwork}
\end{figure*}

\vspace{12 pt}
\centerline{\textbf{\Large A Unified Framework}}
In nonconvex optimization problems, the locally optimal solution is hard to guarantee in the conventional AO framework, and thus monotonic convergence theorem is widely used to prove the convergence of the objective function values \cite{Wu SDR,Huang}. In \cite{Liu}, a new structure is proposed for dealing with nonconvex constraints with no requirement for convex subset of the feasible area. Motivated by this structure, we propose a unified framework named constrained SCA algorithm to find a stationary point for a general nonconvex but smooth optimization problem in RIS-aided networks.

The proposed unified framework is shown in Fig. \ref{Unified_Framwork}. The algorithm is based on solving a sequence of strongly convex objective/feasibility optimization problems obtained via approximating the objective/constraint functions in the original problems. It is worth noting that the problem must satisfy the following two assumptions: i) all functions are continuously differentiable; ii) their derivatives and second-order derivatives are bounded. Besides, there are three assumptions for approximate functions: i) the approximate function is strongly convex and Lipshitz continuous; ii) their derivatives and second-order derivatives are bounded; iii) gradient consistency and value consistency at current solution. These assumptions are quite standard and are readily satisfied for the functions with continuous phase shifts. Hence, when constructing an approximate function, we mainly pay attention to three aspects: strong convexity, gradient consistency, and value consistency. Besides, to execute the proposed algorithm, a diminishing stepsize is adopted in practice. The convergence of the framework is easy to prove by setting the variance of random variables as $\mathbf 0$ in CSSCA \cite{Liu}. In order to distributively solve approximate problem, we can decompose the approximate strongly convex problem into distributively solvable subproblems which are then coordinated by a high-level master problem. Different from centralized implementation, it naturally leads to distributed and parallelizable implementations for a large class of nonconvex problems, especially when there exist a large number of variables to optimize \cite{Scutari}.

In this framework, there exist some distinct advantages. Compared with AO frameworks, the constrained SCA ensures that every limit point satisfying the Slater's condition is a stationary point almost surely. Reference \cite{Schober Large} provides a low-complexity algorithmic framework incorporating AO and gradient-based methods for the problems without coupling in constraints and our proposed framework could be used not only in these problems but also in the problems with coupling in constraints. Compared with convex subset, there are no requirements for any bound, which reduces the burden of handling some sophisticated functions. 

Now, we give an example for the execution of our proposed framework. Considering a general function $f\left(\mathbf w,\boldsymbol{\phi}\right)$, we firstly take $f\left(\mathbf w,\boldsymbol{\phi}^t\right)+f\left(\mathbf w^t,\boldsymbol{\phi}\right)-f\left(\mathbf w^t,\boldsymbol{\phi}^t\right)$ to decouple the conventional variable $\mathbf w$ and phase shift vector $\boldsymbol{\phi}$ at iteration $t$. It is easy to check they have the same gradient and value at $\left(\mathbf w^t,\boldsymbol{\phi}^t\right)$. Then in order to satisfy the strong convexity, we can convexify the nonconvex part via partial linearization. Besides, we add a proximal-like regularization term, in order to relax the convergence conditions of the resulting algorithm or enhance the convergence speed. The decoupling function in the first stage is similar to the method optimizing one variable when the others are fixed. Partial linearization in the second stage is a common approach in SCA algorithms. So it is reasonable to predict that the proposed algorithm enjoys low computational complexity and great convergence.

Besides, in the discrete phase shift system, we are inclined to convert the discrete variable to the continuous one by taking the 6th form $\|\boldsymbol{\phi}\|^2\geq N$ (or $|\phi_n|\geq 1$), $\phi_n\in \mathcal{U}$. The strategy for dealing with the intractable constraint, $\|\boldsymbol{\phi}\|^2\geq N$, is to approximate a challenging problem by an easy-to-handle one via relaxing the constraint to $N - \|\boldsymbol\phi\|^2\leq C$ with some constant $C$ or imposing a penalty on the objective function. Consequently, we can leverage the constrained SCA to cope with the approximate continuous problem.

\begin{figure}[ht]
	\centering
	\fbox{\includegraphics[scale=0.8]{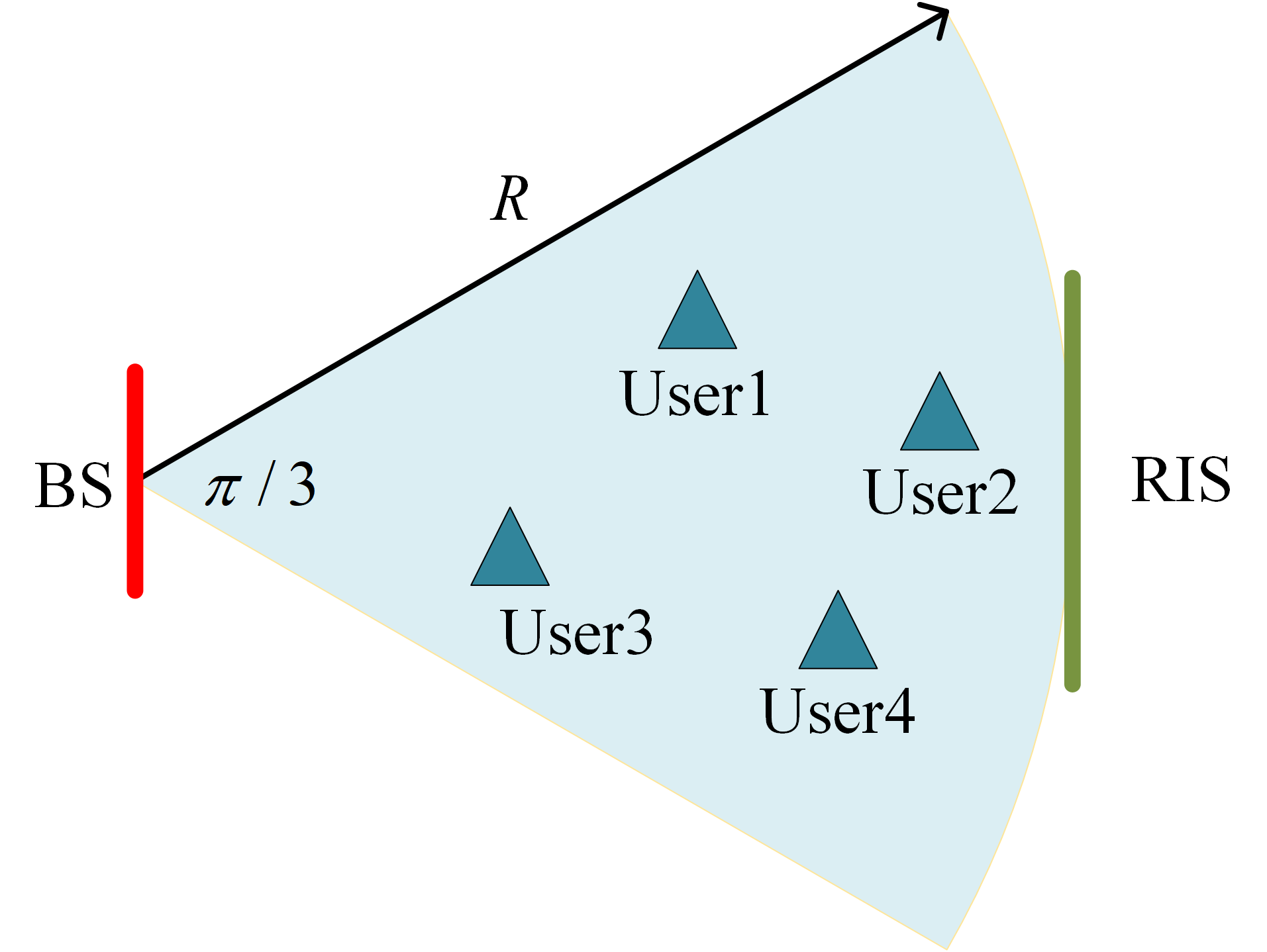}}
	\caption{System model in Power Control}
	\label{Example_system}
\end{figure}

\vspace{12 pt}
\centerline{\textbf{\Large Numerical Results}}
We consider a BS with $N_t=10$ antennas, a RIS with $N=80$ elements, and $K=5$ single-antenna users, with their locations shown in Fig. \ref{Example_system}. All users are randomly and uniformly distributed in the sector and the RIS is deployed at the edge of the cell. It is assumed that both BS-RIS and RIS-User channels are Rician fading with path loss exponent of $2.1$ and Rician factor of $1$, whereas BS-User channel is assumed to follow Rayleigh fading with path loss exponent of $4$. The receiver noise power is $-90$dBm. 

\begin{figure}[ht]
	\centering 
	\subfigure[Convergence of constrained SCA]{
		\includegraphics[scale=0.6]{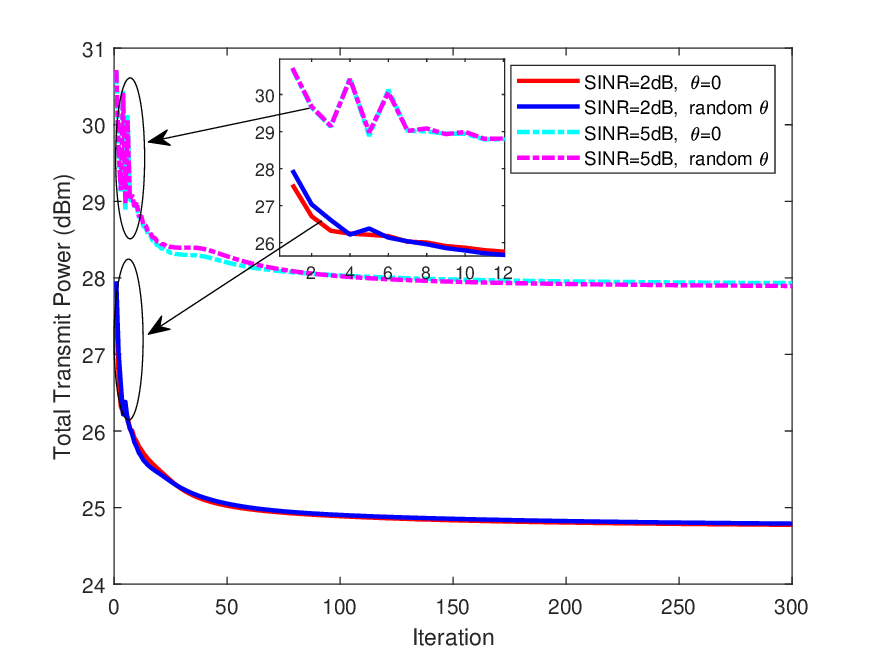} \label{Sim1}
	} 
	\subfigure[Performance comparison]{
		\includegraphics[scale=0.6]{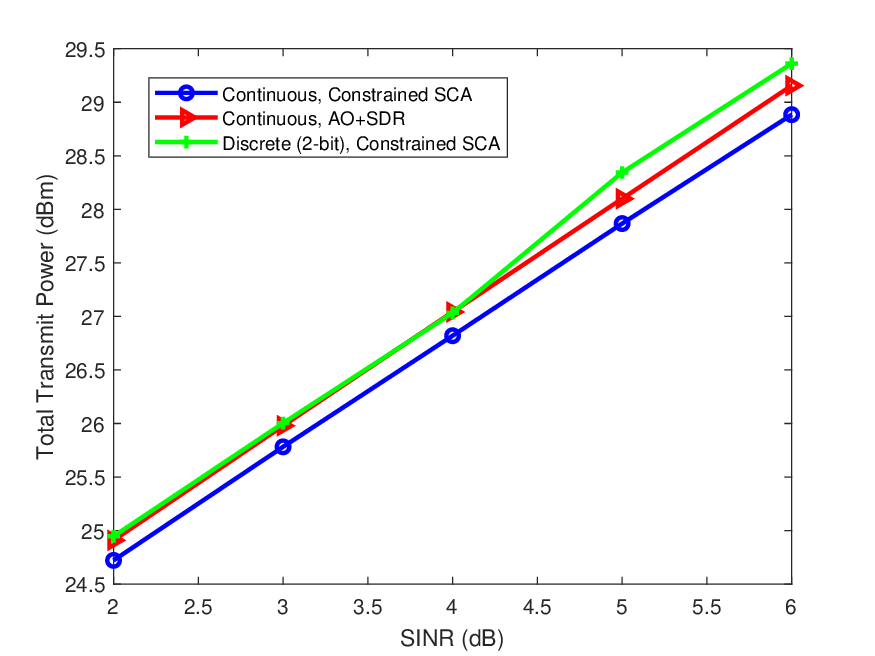} \label{Sim2}
	} 
	\caption{Simulations in Power Control}
	\label{Simulations}
\end{figure}

To verify the performance and efficacy of our proposed algorithm, we minimize the power consumption by jointly optimizing the transmit beamforming at the BS and continuous phase shifts at the RIS, subject to individual SINR constraints at all users. In Fig. \ref{Sim1}, we simulate the convergence of the proposed constrained SCA algorithm for different initial points and SINR requirements. Fig. \ref{Sim1} sketches the number of iterations versus the transmit power by considering four cases with configuration given by: case 1: $\mathrm{SINR}=2\mathrm{dB}, \theta_n=0,\forall n$; case 2: $\mathrm{SINR}=2\mathrm{dB}$, random $\theta_n$; case 3: $\mathrm{SINR}=5\mathrm{dB}, \theta_n=0,\forall n$; case 4: $\mathrm{SINR}=5\mathrm{dB}$, random $\theta_n$. As can be observed, the curve is not always monotonic decreasing because the generated point at each iteration does not always lie in the feasible area, but the algorithm can gradually adjust approximate functions to meet the requirement of SINR. As we can see, the constrained SCA always converges to the same value with the same targeted SINR and the difference between them is almost negligible in all the considered cases. Simulation results illustrate that the proposed algorithm is accurate to optimize the optimization problem in RIS-networks. On the other hand, to demonstrate the effectiveness of the proposed constrained SCA, we adopt the state-of-the-art AO+SDR algorithm as a benchmark. As can be seen from Fig. \ref{Sim2}, our constrained algorithm significantly outperforms AO+SDR algorithm at any targeted SINR from 2 dB to 6 dB. Compared to our proposed scheme, the performance loss of AO+SDR algorithm is mainly due to no guarantee of a stationary point. Besides, the design of discrete phase shifts is also shown in Fig. \ref{Sim2}. As expected, adopting 2-bit RIS phase shifts sacrifices a little performance.

\vspace{12 pt}
\centerline{\textbf{\Large Challenges and Open Problems}}
There are many challenges and issues that need to be addressed for optimization techniques. Some of these challenges are discussed in the following.

\textbf{Large-scale Optimization}: Large-scale optimization problem always emerges in the future communication networks. Since RIS has the advantage of low cost, many works prefer to introduce massive low-cost passive reflecting elements in a RIS or multiple RISs. Besides, the future networks will take massive MIMO technology to accommodate a massive number of devices. As a result, there will exist a large number of variables to optimize. Therefore, how to design an effective large-scale optimization program will be a major challenge.

%
%

\textbf{Finite Resolution Systems}: In practice, it is worthwhile to study the finite resolution systems since hardware limitations make the infinite resolution systems hard to implement. The finite resolution mainly refers to discrete phase shifts, a finite number of controllable amplitudes, etc. The optimization of the phase shifts, the amplitudes, and the number of required quantization levels poses new challenges for the system design.

\textbf{Machine Learning}: Supervised learning, unsupervised learning, and deep reinforcement learning have potentials to solve complex optimization problems. The in-depth survey of machine learning techniques is of great importance and interest in the RIS research field.
		

\vspace{12 pt}
\centerline{\textbf{\Large Conclusion}}
RIS-aided networks have the potential to significantly improve the energy/spectral efficiency and increase the amount of user connectivity. In order to enable RIS techniques to be practically applied to wireless communication, it is imperative to understand and tackle the challenges associated with it. This article surveys the latest research results about the optimization in RIS-aided networks. Firstly, we categorize most optimization problems into the four basic optimization types. Then, we emphasize some common schemes to cope with two key issues, namely phase shift form and decoupling methods. Furthermore, in classic AO framework, we introduce in detail how to exploit the aforementioned technologies flexibly. In RIS-aided networks with continuous phase shifts, we propose a unified framework for the generated subsequence to converge to a stationary point of the optimization problem almost surely. Finally, we clarify the challenges and discuss the open issues as the future research directions.


\begin{thebibliography}{1}
\bibitem{Wu SDR}
Q. Wu, and R. Zhang, ``Intelligent reflecting surface enhanced wireless network via joint active and passive beamforming,'' \emph{IEEE Trans. Wireless Commun.}, vol. 18, no. 11, pp. 5394--5409, Nov. 2019.
	
\bibitem{Huang}
C. Huang, A. Zappone, G. C. Alexandropoulos, M. Debbah and C. Yuen, ``Reconfigurable intelligent surfaces for energy efficiency in wireless communication,'' \emph{IEEE Trans. Wireless Commun.}, vol. 18, no. 8, pp. 4157--4170, Aug. 2019.

\bibitem{Schober IA}
X. Yu, D. Xu, D. W. K. Ng, and R. Schober, ``IRS-assisted green communication systems: Provable convergence and robust optimization,'' 2020, arXiv:2011.06484. [Online]. Available: https://arxiv.org/abs/2011.06484.

\bibitem{Scutari}
G. Scutari, F. Facchinei, and L. Lampariello, ``Parallel and distributed methods for constrained
nonconvex optimization—part I: Theory,'' \emph{IEEE Trans. Signal Process.}, vol. 65, no. 8, pp. 1929--1944, Apr. 2017.

\bibitem{Liu}
A. Liu, V. K. N. Lau, and B. Kananian, ``Stochastic successive convex approximation for non-convex constrained stochastic optimization,'' \emph{IEEE Trans. Signal Process.}, vol. 67, pp. 4189--4203, Aug. 2019.

\bibitem{Schober manifold}
D. Xu, X. Yu, Y. Sun, D. W. K. Ng, and R. Schober, ``Resource allocation for secure IRS-assisted multiuser MISO systems,'' in \emph{Proc. IEEE Globecom Workshops (GC Wkshps)}, Waikoloa, HI, USA, Dec. 2019, pp. 1-6.

\bibitem{Zhaohui}
Z. Yang, M. Chen, W. Saad, W. Xu, M. Shikh-Bahaei, H. Vincent Poor, and S. Cui, ``Energy-efficient wireless communications with distributed reconfigurable intelligent surfaces,'' 2020, arXiv:2005.00269. [Online]. Available: https://arxiv.org/abs/2005.00269.

\bibitem{Wu Discrete}
Q. Wu, and R. Zhang, ``Beamforming optimization for wireless network aided by intelligent reflecting surface with discrete phase shifts,'' \emph{IEEE Trans. Commun.}, vol. 68, no. 3, pp. 1838--1851, Mar. 2020.

\bibitem{Abeywickrama Practical}
S. Abeywickrama, R. Zhang, Q. Wu, and C. Yuen, ``Intelligent reflecting surface: Practical phase shift model and beamforming optimization,'' \emph{IEEE Trans. Commun.}, vol. 68, no. 9, pp. 5849--5863, Sept. 2020.
		
\bibitem{Li Practical}
H. Li, W. Cai, Y. Liu, M. Li, Q. Liu, and Q. Wu, ``Intelligent reflecting surface enhanced wideband MIMO-OFDM communications: From practical model to reflection optimization,'' \emph{IEEE Trans. Commun.}, Mar. 2021, [Early Access].

\bibitem{Schober Large}
Y. Ma, Y. Shen, X. Yu, J. Zhang, S. H. Song, and K. B. Letaief, ``A low-complexity algorithmic framework for large-scale IRS-assisted wireless systems,'' in \emph{Proc. IEEE Globecom Workshops (GC Wkshps)}, Taipei, Taiwan, Dec. 2020, pp. 1-6.

\bibitem{You Penalty}
L. You, J. Xiong, D. W. K. Ng, C. Yuen, W. Wang, and X. Gao, ``Energy efficiency and spectral efficiency tradeoff in RIS-aided multiuser MIMO uplink transmission,'' \emph{IEEE Trans. Signal Process.}, vol. 69, pp. 1407--1421, 2021.

\bibitem{Ming-Min PDD}
M.-M. Zhao, Q. Wu, M.-J. Zhao, and R. Zhang, ``Intelligent reflecting surface enhanced wireless network: Two-timescale beamforming optimization,'' \emph{IEEE Trans. Wireless Commun.}, vol. 20, no. 1, pp. 2--17, Jan. 2021.

\bibitem{Gui KKT}
G. Zhou, C. Pan, H. Ren, K. Wang, and A. Nallanathan, ``Intelligent reflecting surface aided multigroup multicast MISO communication systems,'' \emph{IEEE Trans. Signal Process.}, vol. 68, pp. 3236--3251, 2020.

\bibitem{Huayan FP}
H. Guo, Y.-C. Liang, J. Chen, and E. G. Larsson, ``Weighted sum-rate maximization for reconfigurable intelligent surface aided wireless networks,'' \emph{IEEE Trans. Wireless Commun.}, vol. 19, no. 5, pp. 3064--3076, May 2020.
\end{thebibliography}
\end{document}